\documentstyle[12pt]{article}
\begin{document}
\tolerance=5000
\def\cL{{\cal L}}
\def\cN{{\cal N}}
\def\be{\begin{equation}}
\def\ee{\end{equation}}
\def\bea{\begin{eqnarray}}
\def\eea{\end{eqnarray}}
\def\nn{\nonumber \\}
\def\Y{\bar Y}
\def\z{\zeta}
\def\Z{\bar \zeta}
\vskip -2cm
\ \hfill
\begin{minipage}{3.5cm}
hep-th/9908198\\
August 1999 \\
\end{minipage}
\vfill
\begin{center}
{\large \bf Dilatonic AdS-Kerr Solution to AdS/CFT\\
\medskip
Correspondence}
\vfill
{\bf Alexander BURINSKII}\footnote{ e-mail : grg@ibrae.ac.ru}
\vfill
{NSI Russian Academy of Science, \\
B. Tulskaya 52, Moscow 113191, RUSSIA}
\vfill
{\bf Abstract}
\end{center}
We consider the $ AdS_5$ solution deformed by a
non-constant dilaton interpolating between the standard AdS (UV region) and
flat boundary background (IR region). We show that this dilatonic solution
can be generalized to the case of a non-flat boundaries provided that the
metric of the boundaries satisfies the vacuum Einstein field equations.
\par
As an example, we describe the  case when the four-dimensional boundaries
represent the Kerr space-time.
\bigskip
\newpage
The recently discovered recently remarkable AdS/CFT correspondence between
higher-dimensional classical supergravity  and quantum gauge theory on
the boundary (bulk/boundary correspondence) \cite{WMG} gives  a new
insight to the understanding of strongly coupled gauge theories.
In particular, the $AdS_5 \times S^5$ vacua of type-IIB supergravity
correspond to the four-dimensional Super Yang Mills theory on the
boundary. This stimulated attempts to get the description of running gauge
coupling of Yang-Mills theory and QCD-confinement in the frames of type-0
superstring theory \cite{KTs,Min,FerMar}.
In another approach \cite{KSf,KSf1,LTs,KPR,NO}, the attempts to reproduce
similar QCD-effects were based on the non-supersymmetric background solutions
of type-IIB string theory which can be obtained due to the deformation of
$ AdS_5\times S^5$ vacuum by a non-constant dilaton breaking the
supersymmetry and conformal symmetry. The exact solution of a such sort
was first given in \cite{NO1} and we shall follow the notations of this
work.
Starting with ten-dimensional dilatonic gravity and the solution with
topology of $AdS_5 \times S^5$, one can integrate out five coordinates of
the sphere $S^5$ and obtain \cite{NO1,NO} effective action for $AdS_5$
background
\be \label{i} S=-\int d^5 x \sqrt{-g}\left(R - \Lambda
- {1 \over 2} g^{\mu\nu}\partial_\mu \phi \partial_\nu \phi \right)\ ,\ee
and the equations for metric $g$ and dilaton field $\phi$
\be \label{ii}
R_{\mu\nu}-{1 \over 2}g_{\mu\nu}R + {\Lambda \over 2}g_{\mu\nu}
-{1 \over 2} \left(\partial_\mu\phi\partial_\nu\phi
-{1 \over 2}g_{\mu\nu}g^{\rho\sigma}\partial_\rho \phi
\partial_\sigma \phi \right) =0,
\ee
\be \label{iii}
\partial_\mu\left(\sqrt{-g}g^{\mu\nu}\partial_\nu\phi\right)=0 \ .\ee
Assuming that solutions for $g$ and $\phi$ depend
only on coordinate $y\equiv x^5$, the following ansatz for metric
was proposed
\be \label{v}ds^2=\sum_{\mu,\nu=0}^d g_{\mu\nu}dx^\mu dx^\nu
=f(y)dy^2 + g(y)\sum_{i,j=0}^{d-1}\eta_{ij}dx^i dx^j , \ee
where $\eta_{ij}$ is the metric of Minkowski background.
As it was shown in \cite{NO1,NO}, the solution of these equations can be
given in five dimensions by functions
\be\label{xi}g=y\, \ee
and
\be\label{x} f={3 \over y^2\left(
\lambda^2 + { c^2 \over {2 y^4}}\right)}\ ,\ee
where $\lambda^2 =-\Lambda$ is positive.
The main peculiarity of this solution is the appearance of two boundaries
corresponding to different limiting values of the dilaton.
This solution interpolates between conformal AdS
background (weak coupling regime, $y\rightarrow \infty$) and flat space
at singular value of dilaton $ \phi \rightarrow 6^{1\over 2} \ln y$
(strong coupling regime, $y\rightarrow 0$).
\par
The aim of this note is to attract attention to the possibility of the
generalization of this solution for non-flat four-dimensional
boundaries. We shall show that any metric of the form
\be
\label {hg} ds^2 = g_{\mu \nu} dx^\mu dx^\nu =
f(y) dy^2 + g(y) \hat g_{ik} dx^i dx^k
\ee
represents the solution of dilatonic gravity equations (\ref{ii})
and (\ref{iii}) provided that the boundary metric $\hat
g_{ik}$ satisfies the vacuum Einstein equations. This result
is also valid for a d-dimensional boundary of $AdS_{d+1}$.  As an
example we consider in more details d=4 case when the four-dimensional
boundary represents the Kerr geometry.
\par
Let us consider d+1-dimensional metric (\ref{hg})
where $\hat g_{ik}$ is a d-dimensional boundary metric,
$\mu,\nu...=0,1,...d$, and $i,k,...=0,1...d-1$.
In  the matrix notations we have
\begin{equation}
g_{\mu \nu}=
\left( \begin{array}{cc}
g(y) \hat g_{ik}&0 \\
0&f
\end{array} \right) ,
\end{equation}
while the contravariant form of this metric is given by
\begin{equation}
g^{\mu \nu}=
\left( \begin{array}{cc}
g(y)^{-1} \hat g^{ik}&0 \\
0&f^{-1}
\end{array} \right) ,
\end{equation}
where $ \hat g^{ik} $ is the contravariant form of the
corresponding d-dimensional metric.
  The expressions for the connection
 coefficients $\Gamma^\mu _{\nu \lambda}$ are given in Appendix.
For the
Ricci curvature tensor we have \begin{eqnarray} R_{dd}&= -d (\frac {g''}{2g}
+ (\frac {g'}{2g})^2 +\frac {f'g'}{4fg},\\ R_{id}&= 0, \\ R_{ik}&= \hat
R_{ik} + g_{ik} \lbrack \frac {f'g'}{4f^2}-\frac {g''}{2f} + (2-d)\frac
{(g')^2}{4fg}\rbrack .  \end{eqnarray}
The important point is that  $ \hat R_{ik}$ represents
Ricci curvature of d-dimensional boundary with the metric
$\hat g_{ik}$ and the connections $\hat \Gamma ^l_{ik}$.  Similarly, the
scalar curvature is \begin{equation} R=g^{-1} \hat R  + \frac{d}{4f} \lbrack
2 \frac {f'g'}{fg}-4\frac {g''}{g} + (3-d)(\frac {g'}{g})^2 \rbrack .
\end{equation}
Finally, the expressions for the Einstein  tensor
are given by
$G_{\mu\nu}= R_{\mu\nu} - \frac{1}{2} g_{mu\nu}R $ and are as follows:
\begin{eqnarray} G_{dd} &= \frac{d(d-1)}{8} (\frac{g'}{g})^2 , \\ G_{id}
&=0,\\ G_{ik}&=\hat G_{ik} + g_{ik} \lbrack \frac{(d-1)(d-4)(g')^2}{8fg} -
 \frac {(d-1)f'g'}{4f^2} +\frac {(d-1)g''}{2f}\rbrack. \end{eqnarray}
When the d-dimensional metric $\hat g_{ik}$ satisfies the vacuum Einstein
field equations $\hat G_{ik}=0$, the components  $\hat G_{ik}$ drop out of
the equations (\ref{ii}) and (\ref{iii}).
As a result we come exactly to differential equations for functions
$f$ and $g$ given in the paper \cite{NO} and to the expressions
(\ref{xi}) and (\ref{x}).  Therefore, in the solutions (\ref{v}),
(\ref{xi}) and (\ref{x}) the d-boundary metric $\eta _{ik}$ can
be replaced by any metric $\hat g_{ik}$ satisfying the vacuum Einstein
equations.
\par
As an example let us now consider the partial case $d=4$ and the Kerr
boundary metric $\hat g_{ik}$ in the Kerr-Schild form
\begin{equation}
\hat g_{ik} =
\eta _{ik} + 2 h k_{i} k_{k}  ,
 \label{2}
\end{equation}
where $h$ is  the harmonic scalar function
\begin{equation}
h = m r/(r^2+a^2 \cos ^2 \theta),
 \label{3}
\end{equation}
and $m$ is the mass parameter.
This is  an important solution of the vacuum Einstein equations describing
the field of rotating black holes and modelling the gravitational field
of spinning particle \cite{part,source}. Besides, this form allows one to get
a simple comparison with the Minkowskian case.
\par
The vector field $k_i (x)$ is tangent to the principal null congruence and
it is determined by the 1-form
\begin{equation}
k_i dx^i = \frac{\sqrt{2}}{1+Y \Y} \lbrack
du+ \Y d \z  + Y d \Z - Y \Y d v \rbrack,
 \label{4}
\end{equation}
where
\begin{eqnarray}
2^{1\over 2}\z &=& x+iy ,\qquad 2^{1\over 2 } \Z = x-iy ,\nonumber \\
2^{1\over 2}u &=& z + t ,\qquad 2^{1\over 2}v = z - t ,
\label{5}
\end{eqnarray}
are the Cartesian null coordinates,
and $Y(x)=e^{i\phi} \tan {\theta \over 2}$ is the projective angular
coordinate.  \footnote{It can be also expressed as $Y= (z-ia -\tilde
r)/(x-iy)$, where $\tilde r =r +i a \cos \theta$ is complex radial distance.}
The field $k_i$ is null in respect to Minkowski metric,
$k_i k_k \eta ^{ik}=0$, as well as regarding the metric $\hat g_{ik}$
\be \label{null}
k_i k_k \hat g^{ik}=0.
\ee
One can build the five dimensional field $ k^\mu = (k^i,0)$ which  will
obviously be the null field with respect to the full five-dimensional
metric as well \be \label{5null} k^\mu k^\nu g_{\mu\nu}=0.\ee
Therefore, the five-dimensional Kerr-AdS metric takes the form \be \label
{KADS} ds^2 = f(y) dy^2 + g(y) [\eta _{ik} +2h k_i k_k] dx^i dx^k, \ee
where functions $f(y)$, $g(y)$ and $h$ are given by (\ref{x}),
(\ref{xi}) and (\ref{3}), and vector field $k_i$ is determined by expressions
(\ref{4}) and (\ref{5}).
\par  In the region of parameters
corresponding to spinning particles the black hole horizons disappear and a
region of the rotating disk-like source is opened. The structure of this
source should possess some properties of OCD-confinement and represents
an old problem. The predicted exotic properties of the matter of this source
\cite{source} do not allow to construct them in four dimensions from a known
sort of classical matter. The conjectured AdS/CFT correspondence
gives a new approach to this problem and stimulates consideration of new
models, in particular, of a bag-like models resembling the cosmic
bubble models. In this case the bag-like source based on the supersymmetric
domain wall models \cite{bubble} can contain the AdS-region of a false vacuum
inside the bag separated from the true vacuum of the outer region by a thin
domain wall.  As a field model, the supersymmetric version of
$U(1) \times U(\tilde 1)$  Witten model \cite{mor} seems the most appropriate,
since it provides the long range electromagnetic field out of core.
A hypothetical mechanism of the formation of the bag-like Kerr source can be
connected with a phase transition governed by the value of dilaton
near the core.
\par
The AdS-BH solution of another sort was considered in \cite{NO2}.
\par
{\bf Acknowledgement}
\par
I am thankful to Sergei Odintsov for
paying my  attention to this area and for useful discussions.
\section*{Appendix}
\medskip
The connection coefficients are (no summation over $d$)
\begin{eqnarray} \Gamma^d_{dd} &= \frac {f'} {2f}, \\
 \Gamma^d_{id} &= 0, \\
\Gamma^d_{di} &= 0, \\ \Gamma^d_{ik} &= - \frac{g'g_{ik}}{2f}, \\
\Gamma^i_{dk} &= \Gamma^i_{kd} = - \frac{g'\delta^i_k}{2g}, \\ \Gamma^i_{dd}
&= 0, \\
 \Gamma^i_{jk} &= \hat \Gamma^i_{jk}.
 \end{eqnarray}
 Here $\hat \Gamma^i_{jk}$ are connections to d-dimensional metric
 $\hat g_{ik}$.
\par
We have also the relations
\begin{eqnarray} \Gamma^\mu_{d\mu}
&=\Gamma^\mu_{\mu d} = \frac {f'} {2f} + d\frac {g'} {2g},\\
 \Gamma^\mu_{i\mu} &=\Gamma^\mu_{\mu i} =
\hat \Gamma^\mu_{i\mu} =\hat \Gamma^\mu_{\mu i}, \\
\Gamma^\mu_{d\nu}\Gamma^\nu_{d\mu} &=
(\frac {f}{2f})^2 + d(\frac {g'}{2g})^2,\\
\Gamma^\mu_{i\nu}\Gamma^\nu_{k\mu} &=
\hat \Gamma^l_{ij} \hat \Gamma^j_{kl} - g_{ik}\frac {(g')^2}{2gf},\\
\Gamma^\mu_{dd}\Gamma^\nu_{\nu\mu} &=\frac{f'}{2f}
( \frac {f'} {2f} + d\frac {g'} {2g}), \\
\Gamma^\mu_{ik}\Gamma^\nu_{\nu\mu} &=
\hat\Gamma^j_{ik}\hat\Gamma^l_{lj} - g_{ik} \frac{g'}{2f}
( \frac {f'} {2f} + d\frac {g'} {2g}).
\end{eqnarray}

\end{document}